\documentclass[10pt,twocolumn,twoside]{IEEEtran}

\usepackage[dvips]{epsfig}
\usepackage[dvips]{graphicx}
\usepackage{amsmath,amsfonts,bm,amssymb,psfrag,ifthen,color,subfigure}
\usepackage{algpseudocode}
\usepackage{algorithmicx}
\usepackage{setspace}
\usepackage{cite}
\usepackage{longtable}
\usepackage{stfloats}  
\usepackage{graphics,booktabs,color,subfigure}
\usepackage{multirow}
\usepackage{makecell}
\usepackage{amssymb}
\usepackage{footnote}
\makesavenoteenv{table}

\begin{document}

\title{Signal Demodulation with Machine Learning Methods for Physical Layer Visible Light Communications:  Prototype Platform, Open Dataset and Algorithms}
 \author{Shuai~Ma, Jiahui Dai,  Songtao Lu, Hang Li,  Han Zhang,  Chun Du,    and  Shiyin Li

\thanks{Manuscript received December 15, 2018;   }
\thanks{S. Ma  is with the School of Information and Control   Engineering, China
University of Mining and Technology, Xuzhou 221116,
China, and also with the State Key Laboratory of Integrated Services Networks, Xidian University, Xi'an 710071, China
 (e-mail: mashuai001@cumt.edu.cn).}
\thanks{ J. Dai, C. Du,   and S. Li   are with the School of Information and Control   Engineering, China
University of Mining and Technology, Xuzhou 221116,
 China (e-mails: daijiahui@cumt.edu.cn;
 duchun@cumt.edu.cn;  lishiyin@cumt.edu.cn).}
\thanks{S. Lu is with the Department of Electrical and Computer Engineering, University of Minnesota, Minneapolis, MN 55455, USA (e-mail: lus@umn.edu).}
\thanks{H. Li  is with the Shenzhen Research Institute of Big Data, Shenzhen 518172, China (email: hangdavidli@163.com).}
\thanks{H. Zhang is with the Department of Electrical and Computer Engineering, University of California, Davis, CA
95616, USA (e-mail: hanzh@ucdavis.edu).}
}
\maketitle

\begin{abstract}
In this paper, we investigate the design and implementation of  machine learning (ML) based demodulation methods in the physical layer of visible light communication (VLC) systems.
 We build
   a flexible hardware prototype of an end-to-end VLC system, from which the received signals are collected as the real data.
   The dataset is available online,
  which contains  eight types of modulated  signals.
  Then, we propose three  ML demodulators based on convolutional neural network (CNN), deep belief network (DBN),
 and adaptive boosting (AdaBoost), respectively. Specifically,   the CNN based demodulator  converts  the  modulated signals to  images and recognizes the signals by the  image classification. The proposed DBN based demodulator contains three  restricted Boltzmann machines (RBMs) to extract the modulation features.
 The AdaBoost method includes a strong classifier that is constructed by  the weak classifiers with the $k$-nearest neighbor (KNN) algorithm.
  These three demodulators are trained and tested by our online open  dataset.
     Experimental results show that the  demodulation accuracy of the three data-driven demodulators  drops as the transmission   distance increases.
  A higher modulation order negatively influences the accuracy for a given transmission distance.
Among the three ML methods, the AdaBoost modulator achieves the best performance.
  \end{abstract}


\section{Introduction} \label{sec:intro}

With the rapidly increasing number of mobile
digital devices and the soaring high volume of wireless data traffic, the  high speed wireless transmission is also highly demanded.
Traditional radio frequency
(RF)   systems  are currently facing spectrum crisis, which is the bottleneck of enhancing the network
capacity  \cite{Chandrasekhar}.
Visible light communication (VLC),  with advantages like huge unregulated spectrum, high security  and immunity to electromagnetic interference,
  has sparked significant   research attention as a
promising solution for short range wireless communications \cite{IEEE}.
  Through massive deployment of light-emitting diodes (LEDs), VLC  typically employs the intensity modulation and direct
detection (IM/DD) technique  for both the illumination and data transmissions \cite{Komine,Elgala,Arnon,Jovicic,Pathak,MultiUser,RateMaximized},
where the signal is recovered by capturing  fluctuations of optical intensity.

Demodulation  of radio signals
plays a  fundamental role  in VLC systems.
In general, the traditional demodulators could be categorized into two classes:  coherent and non-coherent demodulators.
Moreover, the priori knowledge, such as  channel state information (CSI) or channel noise, is usually required.
Most of previous works \cite{Fath,Wang,Ying} indeed assume
that each receiver can accurately estimate the fading coefficients.
 In slow-fading scenarios,
such CSI might be obtained via estimation from training
sequences. However, in fast-fading
scenarios, CSI is usually hard to estimate since the fading
coefficients vary quickly within the period of one transmission
block.
    Besides,  most of existing works assume that the VLC channel suffers from additive white Gaussian noise (AWGN), and thus the applied demodulators are optimal in terms of
the AWGN  channel.
 However, the practical VLC channels are not easy to model since there exist too many factors, including but not limited to: limited modulation bandwidth of LEDs,
 multi-path dispersion, impulse noise, spurious
or continuous jamming, and low
sensitivity of commercial photodetector (PD). Even though the channel can be approximated by a complex model, the non-casual  knowledge of the channel model might not be
available at the receiver, especially when the channel fading is non-stationary
   with unknown distributions.

Given the above issues, machine learning (ML) based model-free demodulators
become more attractive, where the requirements for
the priori knowledge can be widely relaxed or even removed \cite{Mitchell2003Machine}.
 In \cite{ChannelEstimation,DownlinkMIMO,dlNOMA}, the authors used neural networks which are considered as black boxes to detect the channel condition but with high computational complexity.
Since the information of the  modulated signals  is represented by the
 amplitude and phase, feature extraction is critically important to the signal demodulation.
Note that in conventional RF systems, ML based demodulators have been investigated, such as   a neural network demodulator \cite{Mursel} and  a one-dimensional convolutional neural network (CNN) based demodulator \cite{Zhang2018Enhanced} for
   binary phase shift keying (BPSK) signals.
Also, a deep convolutional neural  network (DCNN) demodulator
 was proposed in \cite{mixedsignals} to  respectively demodulate symbol sequences from mixed signals.
   In \cite{DemodulatorbasedDBN}, the authors showed that  the deep belief network (DBN) based
demodulator is feasible for the AWGN channel with certain channel impulse
response and the Rayleigh non-frequency-selective flat fading
channel.
In \cite{Lanting},
   a deep learning (DL) based detection method was proposed for signal demodulation in short range multipath
channel without any channel equalization.

Different from RF communication
   systems,  the transmitted  signals of VLC  should be  real and non-negative due to the IM/DD mechanism.
   In the research of VLC systems,
   the  ML based approaches have been  investigated to some extent.
In \cite{Lee2018Deep},  a DL based  autoencoder  was  designed   for  multi-dimensional color modulation in multicolored VLC systems, which   can reduce the    average symbol error probability.
A soft binarization  training strategy was proposed for   autoencoder VLC systems in \cite{Lee2018Binary}, which
   yielded an efficient on-off keying (OOK)
transceiver over general optical channels.
However,   the existing works\cite{Mursel,Lanting,DemodulatorbasedDBN,mixedsignals,Zhang2018Enhanced
,Lee2018Deep,Lee2018Binary} are based on synthetic data
  rather than real datasets.
To the best of our knowledge, the ML based demodulation schemes have not been well studied in VLC systems, and there is no open real measurement data yet.

  In this paper, we present a unified data-driven framework of demodulation  by
 ML approaches.  To be specific,
we propose three data-driven demodulation methods: CNN, DBN,
adaptive boosting (AdaBoost) \cite{Freund1996Experiments}  based demodulators for end-to-end VLC systems.
 Also,  the performance of the  three  data-driven demodulators are evaluated for
 the  different  modulation schemes  via the real measured data.
Our main contributions
are as follows:
\begin{itemize}

 \item We
  propose a flexible end-to-end VLC system hardware
prototype  to study data-driven  demodulation approaches.
 By exploiting this prototype, we collect received signal data in real physical environments in eight modulation schemes, i.e., OOK, quadrature phase shift keying (QPSK), 4-pulse position modulation (PPM), 16-quadrature amplitude modulation (QAM),
32-QAM, 64-QAM, 128-QAM and 256-QAM. We establish an open online real modulated  dataset available at
https://pan.baidu.com/s/1rS143bEDaOTEiCneXE67dg,
 where the transmission distance  of the  eight modulated  signals is measured from $0$cm to $140$cm.
 To the best of our knowledge,
this is the first open real modulated signals dataset of VLC systems.

\item Three ML-based demodulators are designed. We propose a CNN based demodulator with
two convolutional layers  and two pooling layers. It first converts the  modulated signals to
images and then identifies the signals by the  image classification.
Then we develop a DBN based demodulator with  three restricted Boltzmann machines (RBMs) to extract the modulation features.
Finally, an AdaBoost based demodulator  is presented, where a strong classifier is constructed by  several weak classifiers with the $k$-nearest neighbor (KNN) algorithm.

\item Based on the established real dataset,  we investigate the demodulation performance of the proposed  three data-driven demodulators.
 Specifically, the  demodulation accuracy of the three ML based demodulators is decreasing over the transmission and the modulation order for a fixed transmission distance.
     Experimental results also show that the demodulation accuracy of the AdaBoost based demodulators is higher than other demodulators.
Moreover, for the  short distance or high SNR scenario,
a high-order modulation is preferred.

 \end{itemize}

\emph{Notation:} The following notations are used throughout the paper. Bold upper case letters represent matrices, e,g, ${\bf{A}}$.
Bold lower case letters represent vectors, e,g, ${\bf{a}}$.
${[{\rm{\cdot}}]^{\rm{T}}}$ means transpose, and ${\mathop{\rm Re}\nolimits} \left[ {\rm{\cdot}} \right]$ is used to obtain the real part.
${[{\rm{\cdot}}]_{p,q}}$ indicates the element at the $p$th row and the $q$th column.
Moreover, ${[{\rm{\cdot}}]_{p}}$ indicates the  $p$th element.
${\left\| {\rm{\cdot}} \right\|_2}$ is the ${L_2}$ norm operator, and $\mathbb{R}$ is the real number sets.
The natural logarithm $\ln \left( {\rm{\cdot}} \right)$ is used.
$ \approx $ means approximate equals, and  $\sim$ means subjecting to certain distribution.
$\partial$ denotes the partial derivation and $*$ is the convolution operator. $\leftarrow $  means that the values on the left is updated by the values on the right.

\section{System Model}

\begin{figure}[h]
          \centering
      \includegraphics[width=9cm]{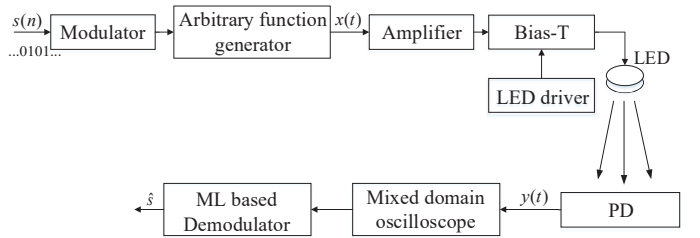}
 \caption{~The VLC system model with ML based demodulator.}
  \label{VLC sys} 
\end{figure}

 As   illustrated in Fig. \ref{VLC sys},
we propose a flexible end-to-end VLC prototype,
which consists a modulation block, an arbitrary function generator, an amplifier, a bias-T, a  LED driver,
 a single LED,   a single PD,
  a mixed domain oscilloscope, and a ML based demodulation block.
  According to Fig. $1$, the digital signal $s\left( n \right)$ is modulated by the $M$-QAM scheme,
  converted to the analog   signal by the    arbitrary function generator, and further amplified  by the amplifier. After amplification, the signal adds
    the   direct current (DC)   at the Bias-T.
  Finally, the signal is transformed to the visible
light by LED, and sent out to the wireless channels.
  At the receiver, the optical signal from LED is converted to  the analog signal   through PD, and then the   analog signal is
   converted to a digital signal  at the mixed domain oscilloscope. Afterwards, the   digital    signal is demodulated by the ML based demodulator.

By exploiting  digital modulation schemes, {such as $M$-QAM and $M$-PPM},
the  transmitted signal $x\left( t \right)$  is given as
\begin{align}
x\left( t \right) = {\mathop{\rm Re}\nolimits} \left[ {s\left( t \right)p\left( t \right){e^{j2\pi {f_c}t}}} \right],~0 \le t \le T,
\end{align}
where $s\left( t \right)$ denotes the  baseband signal,  ${f_c}$
 denotes the carrier  frequency,   ${p\left( t \right)}$ stands for the  signal pulse, and $T$ represents the period of  signal.

Let $g$ denote the channel gain  between  LED and PD, which includes both  line-of-sight (LOS) path and multi-reflection paths.
 At the PD side, the  received signal $y\left( t \right)$ is given as:
\begin{align}
y\left( t \right) = gx\left( t \right) + n\left( t \right),
\end{align}
where $n\left( t \right)$ is the received noise.
 Via the mixed domain oscilloscope , the received analog signal $y\left( t \right)$ is sampled to digital signals.

Let ${{{\bf{\hat y}}}_i} = {[{{\hat y}_{\left( {i - 1} \right)N + 1}},{{\hat y}_{\left( {i - 1} \right)N + 2}},...,{{\hat y}_{iN}}]^{\rm{T}}}$ denote the received signal vector during
the $i$th period,
where ${{\hat y}_{\left( {i - 1} \right)N + n}}$ is
the $n$th sampled point  ${{\hat y}_{\left( {i - 1} \right)N + n}} = y\left( {\frac{{n - 1}}{N}T + \left( {i - 1} \right)T} \right)$, and $N$ is the number of samples during one period.
 Assume that training data set contains $K$ periods of sampled
vectors and $1 \le i \le K$.
Before the demodulator  processing, the received data $\left\{ {{{{\bf{\hat y}}}_i}} \right\}_{i = 1}^K \buildrel \Delta \over = \left\{ {{{\hat y}_1},{{\hat y}_2},...,{{\hat y}_{NK}}} \right\}$ is normalized to $[0,1]$, which can significantly reduce the calculation time of ML \cite{Importance}. The normalized sample ${{\bar y}_i}$ is given by
\begin{align}
{{\bar y}_i} = \frac{{{{\hat y}_i} - {{\hat y}_{\min }}}}{{{{\hat y}_{\max }} - {{\hat y}_{\min }}}},1 \le i \le NK,
\end{align}
where ${{\hat y}_{\min }} = \mathop {\min }\limits_{1 \le i \le NK} {{\hat y}_i}$, ${{\hat y}_{\max }} = \mathop {\max }\limits_{1 \le i \le NK} {{\hat y}_i}$.

 After normalization, we use ${{{\bf{\bar y}}}_i} = {[{{\bar y}_{\left( {i - 1} \right)N + 1}},{{\bar y}_{\left( {i - 1} \right)N + 2}},...,{{\bar y}_{iN}}]^{\rm{T}}}$ to denote the $i$th normalized signal vector.
Moreover,  let ${z_i}$ denote the label for  the  normalized  vector ${{{\bf{\bar y}}}_i}$
and ${\rm{{\cal C}}}$ the label set, i.e., ${z_i} \in {\rm{{\cal C}}}$ for $i = 1,2,...,K$.
The label set ${\rm{{\cal C}}}$ is determined by the modulation scheme.
For example, ${\rm{{\cal C}}} = \{ 1,2,3,4\}$ is used for quadrature phase shift keying QPSK signal.
 Let ${\rm{{\cal T}}} = \left\{ {\left( {{{{\bf{\bar y}}}_1},{z_1}} \right),\left( {{{{\bf{\bar y}}}_2},{z_2}} \right),...,\left( {{{{\bf{\bar y}}}_K},{z_K}} \right)} \right\}$
 denote the   labeled   dataset.

In the following sections, we propose three  ML based demodulators and present their structure in details.

\section{CNN based  Demodulator}

\begin{figure}[h]
          \centering
      \includegraphics[width=9cm]{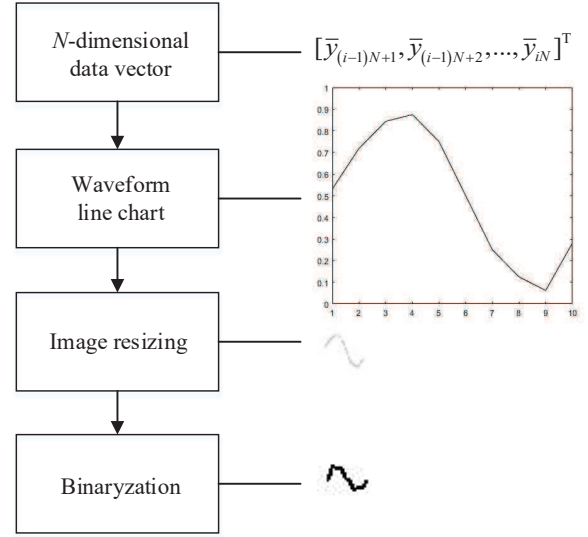}
 \caption{~The  visualization block.}
  \label{cnnpic} 
\end{figure}

Due to  the sparse connectivity and parameter sharing characteristics,  CNN has  a simple structure and strong adaptability and is applied in various domains
  \cite{ModulationClassification,WaterLevel}.
For single carrier modulation, the amplitude and phase information of signal can be extracted for classification.
Therefore, we  investigate  the CNN based demodulator,
which includes a visualization block and a CNN network.
 We first convert
 the data vector ${{\bf{\bar y}}_i}$ into a two-dimensional image format so that the CNN based demodulator can interpret the data as images.
Specifically,  as shown in Fig. \ref{cnnpic},
 the  elements of ${{\bf{\bar y}}_i}$  are first transformed to a point on the two-dimensional plane.
 First, we consider $n$ as the coordinate of horizontal axis and the value of ${{\bar y}_{\left( {i - 1} \right)N + n}}$ as the coordinate of vertical axis,
and transform the vector to $N$ points. Then, we connect these points by polylines,  so that we can obtain the waveform with horizontal axis range of $[1,N]$ and vertical axis range of $[0,1]$.
In waveform images,  both  amplitude  and phase information of the modulated signals are   represented by waveforms with high pixel density.
To reduce the computational load of computer and  preserve the useful information,
 we resize the grey image   with less pixels  by applying the bicubic interpolation algorithm.
Also, the resized grey image is changed into a binary image by the global thresholding algorithm \cite{2007Digital}, which can further distinguish the waveform from the background.
Finally, we obtain  the  output image   matrix ${\bf{X}}$ with a size  $28 \times 28$, i.e, ${\mathbf{X}} \in {\mathbb{R}^{28 \times 28}}$.

\begin{figure}[h]
          \centering
      \includegraphics[width=9cm]{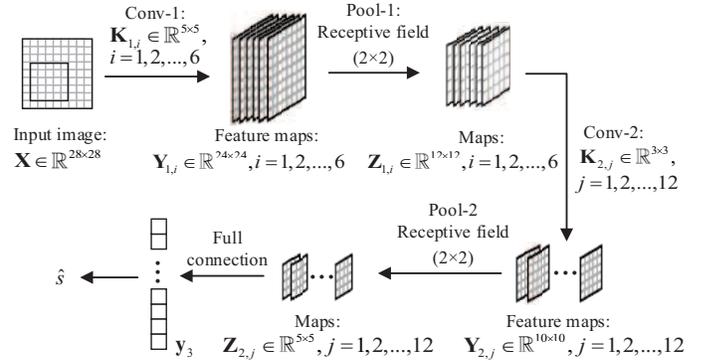}
 \caption{~The structure of CNN.}
  \label{cnn} 
\end{figure}

Then, the output ${\bf{X}}$ of the visualization block  is processed by
  the presented CNN network,
which includes two convolutional layers, two pooling layers, and one full-connected layer as shown in Fig. \ref{cnn}.
Let Conv-$1$ and Conv-$2$ stand for the first and second convolutional layer, respectively.
Moreover, the Pool-$1$ and  Pool-$2$ denote the first and second pooling layer, respectively.

As shown in Fig. \ref{cnn},  the  input image  first convolutes with six
    kernels in Conv-$1$, respectively. Then, the Conv-$1$ outputs six feature maps. In Pool-$1$,  the feature maps
    are compressed to maps by the $\left( {2 \times 2} \right)$ receptive field \cite{ModulationClassification}.
  Then, the maps are processed via kernels, and further compressed to maps by the
    $\left( {2 \times 2} \right)$ receptive field in Pool-$2$.
    Finally,  the output maps of Pool-$2$ are connected via full connection to the output layer , whose
 dimension  is determined by the modulation scheme.

The  parameters  of the CNN are shown in Table \ref{CNNtable}.
  Let ${{\mathbf{K}}_{1,i}}$ represent the $i$th kernel of Conv-$1$,
 ${{\mathbf{K}}_{1,i}} \in {\mathbb{R}^{5 \times 5}}$, $i = 1,2,...,6$.
Moreover, let   ${{\mathbf{Y}}_{1,i}}$  denote the output feature map obtained by ${{\mathbf{K}}_{1,i}}$, which can be expressed by \cite{cnnNotes}
\begin{align}
{{\mathbf{Y}}_{1,i,p,q}} = {\text{sigmoid}}\left( {{b_i} + {{\left[ {{\mathbf{X}}*{{\mathbf{K}}_{1,i}}} \right]}_{p,q}}} \right),
\end{align}
where ${b_i}$ stands for the bias of ${{\mathbf{K}}_{1,i}}$, $p = 1,2,...,24$, $q = 1,2,...,24$.
Here, we choose ${\rm{sigmoid}}\left( x \right) \buildrel \Delta \over = \frac{1}{{1 + {e^{ - x}}}}$ as the activation function.

 \begin{figure}[h]
          \centering
      \includegraphics[width=6cm]{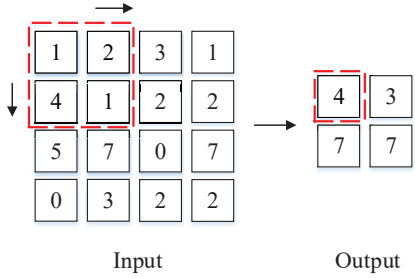}
 \caption{~The max-pooling operation with $2 \times 2$ filter and stride 2.}
  \label{maxp}
\end{figure}

The convolutional layer is followed by Pool-$1$, which is used for down sampling of the output feature maps and increasing the robustness of the model.
The pooling method used in this paper is max-pooling, as   shown in Fig. \ref{maxp}.
The maximum value in a submatrix of size $2 \times 2$ is treated as the local output.
Let ${{\mathbf{Z}}_{1,i}}$ stand for the pooling result of ${{\mathbf{Y}}_{1,i}}$, and it can be expressed by
\begin{align}
{{\mathbf{Z}}_{1,i}} ={\text{pooling}}\left( {{{\mathbf{Y}}_{1,i}}} \right),
\end{align}
where ${\rm{pooling}}\left(  \cdot  \right)$ stands for the component-wise max-pooling function.

Let ${{\mathbf{K}}_{2,j}}$ stand for the kernel adopted in Conv-$2$, ${{\mathbf{K}}_{2,j}} \in {\mathbb{R}^{3 \times 3}}$, $j = 1,2,...,12$.
Assume that ${{\mathbf{Y}}_{2,j}}$ is the output feature map of ${{\mathbf{K}}_{2,j}}$, ${{\mathbf{Y}}_{2,j}} \in {\mathbb{R}^{10 \times 10}}$, $j = 1,2,...,12$. It can be obtained by
\begin{align}
{{\mathbf{Y}}_{2,j,p,q}} = {\text{sigmoid}}\left( {{b_j} + \sum\limits_i {{{\left[ {{{\mathbf{Y}}_{1,i}}*{{\mathbf{K}}_{2,j}}} \right]}_{p,q}}} } \right),
\end{align}
where $p = 1,2,...,10$, $q = 1,2,...,10$.

After Pool-$2$ with receptive field of $2 \times 2$, the output maps ${{\mathbf{Z}}_{2,j}}$ are transformed into a one-dimensional label space by the full-connected layer.
Let ${{{\mathbf{y}}_3}}$ stand for the one-dimensional vector,
%
the output label ${\hat z}$ can be expressed by
\begin{align}
\hat z = \arg \mathop {\max }\limits_i {\left[ {{{\mathbf{y}}_3}} \right]_i}.
\end{align}
The dimension of ${{\mathbf{y}}_3}$ is determined by the modulation scheme employed. Then, the label ${\hat z}$ corresponds to the demodulation result ${\hat s}$.

\begin{table}[htbp]
\caption{Parameters setting of CNN.}
\centering
\begin{tabular}{cccc}
\toprule
Layer & Kernel size & Stride & Output size   \\
\midrule
Input & $\quad$ &  $\quad$  & $28 \times 28$ \\
Conv-1& $5 \times 5$ &  $1$  & $24 \times 24$ \\
Pool-1& $2 \times 2$ &  $2$  & $12\times 12$ \\
Conv-2& $3 \times 3$ &  $1$  & $10 \times 10$ \\
Pool-2& $2 \times 2$ &  $2$  & $5\times 5$ \\
\bottomrule
\end{tabular}
\label{CNNtable}
\end{table}

\section{DBN based Demodulator}
DBN has been widely applied to address many practical problems such as handwritten recognition, speech recognition, and image classification, since it can efficiently extracts high-level and hierarchical
features from the measured signal data by a multiple nonlinear transformation.
RBM is the fundamental block of DBN, which is a realization of undirected
graphical model and contains a layer of visible neurons and a layer of hidden neurons \cite{G2006Reducing}.
It is noted that there are only connections between the visible layer and the hidden layer.
\begin{figure}[h]
          \centering
      \includegraphics[width=7cm]{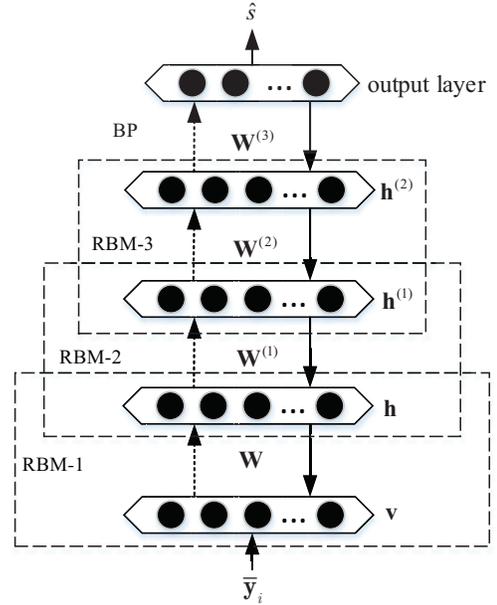}
 \caption{~The structure of DBN.}
  \label{fig:DBN} 
\end{figure}

Consider a DBN  with  three RBMs, as shown in  Fig. \ref{fig:DBN}.
The first RBM is consisted of a visible layer ${\bf{v}} = {\left[ {{v_1},{v_2},...,{v_m}} \right]^{\rm{T}}}$   and a hidden layer ${\bf{h}} = {\left[ {{h_1},{h_2},...,{h_n}} \right]^{\rm{T}}}$, which contains $m$ neurons and $n$ neurons in the visible layer and hidden layer, respectively.
Let ${\bf{W}} = {\left[ {{{\bf{w}}_1},{{\bf{w}}_2},...,{{\bf{w}}_n}} \right]^{\rm{T}}}$ denote the connection weight matrix between ${\bf{v}}$ and ${\bf{h}}$,
 where ${{\bf{w}}_j} = {[{w_{j1}},{w_{j2}},...,{w_{jm}}]^{\rm{T}}}$, $j = 1,2,...,n$.
 Moreover, ${\bf{a}} = {\left[ {{a_1},{a_2},...,{a_m}} \right]^{\rm{T}}}$ and
 ${\bf{b}} = {\left[ {{b_1},{b_2},...,{b_n}} \right]^{\rm{T}}}$ denote the bias of ${\bf{v}}$ and ${\bf{h}}$, respectively.


RBM is an  energy based model, which defines the probability distribution of variables by the energy function.
With the normalized signal ${{{\bf{\bar y}}}_i}$ in ${\rm{{\cal T}}}$, the energy of the first RBM is  given by
\begin{align}
E\left( {{\bf{v}},{\bf{h}}} \right) =  - {{\bf{a}}^{\rm{T}}}{\bf{v}} - {{\bf{b}}^{\rm{T}}}{\bf{h}} - {{\bf{h}}^{\rm{T}}}{\bf{Wv}},
\end{align}
where ${\bf{v}} = {{{\bf{\bar y}}}_i}$.
The probability distribution of the visible layer ${\bf{v}}$ is given by
\begin{align}
p\left( {\bf{v}} \right) = \frac{1}{Z}\sum\limits_{\bf{h}} {{e^{ - E\left( {{\bf{v}},{\bf{h}}} \right)}}},
\end{align}
where $Z = \sum\limits_{{\mathbf{v}},{\mathbf{h}}} {{e^{ - E\left( {{\mathbf{v}},{\mathbf{h}}} \right)}}}$ is a normalization constant.

Then,
the optimal parameters ${\bf{W}},{\bf{a}},{\bf{b}}$  can be obtained by maximizing the log-likelihood function as follows
\begin{align}\label{energy_RBM}
\mathop {\max }\limits_{{\bf{W}},{\bf{a}},{\bf{b}}} \sum\limits_{\{ {{\bf{v}}}\}} {\ln p\left( {\bf{v}} \right)}.
\end{align}
To solve the unconstrained optimization problem \eqref{energy_RBM}, we simply adopt the gradient descent method.
The partial derivations with respect to variables ${\bf{W}}$, ${\bf{a}}$, and ${\bf{b}}$  can be respectively approximated by
\begin{subequations}
\label{derivation}
\begin{align}
\frac{{\partial \ln p\left( {\bf{v}} \right)}}{{\partial {w_{ji}}}} & \approx p\left( {{h_j} = 1\left| {\bf{v}} \right.} \right){v_i} - p\left( {{h_j} = 1\left| {{\bf{\hat v}}} \right.} \right){\hat v_i}, \hfill \\
\frac{{\partial \ln p\left( {\bf{v}} \right)}}{{\partial {a_i}}} &\approx {v_i} - {{\hat v}_i}, \hfill \\
\frac{{\partial \ln p\left( {\bf{v}} \right)}}{{\partial {b_j}}} & \approx p\left( {{h_j} = 1\left| {\bf{v}} \right.} \right) - p\left( {{h_j} = 1\left| {{\bf{\hat v}}} \right.} \right), \hfill
\end{align}
\end{subequations}
where $p\left( {{h_j} = 1\left| {\mathbf{v}} \right.} \right)$ and $p\left( {{h_j} = 1\left| {{\mathbf{\hat v}}} \right.} \right)$ denote the conditional probability distribution of hidden neurons ${\mathbf{h}}$ given ${\bf{v}}$ and ${\bf{\hat v}}$, respectively. ${\bf{\hat v}} = {[{{\hat v}_1},{{\hat v}_2},...,{{\hat v}_m}]^{\rm{T}}}$ denotes the reconstruction of visible states, which can be obtained as follows \cite{CD}.

Given the visible layer ${\bf{v}}$, $p\left( {{h_j} = 1\left| {\mathbf{v}} \right.} \right)$ is given by
\begin{align}\label{h_v}
p\left( {{h_j} = 1\left| {\bf{v}} \right.} \right) = {\rm{sigmoid}}\left( {{b_j} + \sum\limits_{i = 1}^n {{w_{ji}}{v_i}} } \right).
\end{align}

 Then,  we can generate ${\bf{\hat h}} = {[{{\hat h}_1},{{\hat h}_2},...,{{\hat h}_n}]^{\rm{T}}}$ according to distribution \eqref{h_v} as the following:
\begin{align}
{\bf{\hat h}} \sim p\left( {{\bf{h}}\left| {\bf{v}} \right.} \right).
\end{align}

Similarly, the distribution  of the visible layer ${\bf{v}}$  is given by
\begin{align}\label{v_h}
p\left( {{v_i} = 1\left| {{\bf{\hat h}}} \right.} \right) =  {\rm{sigmoid}}\left( {{a_i} + \sum\limits_{j = 1}^m {{{\hat h}_j}{w_{ji}}} } \right).
\end{align}
Then,  the reconstructed data ${{\bf{\hat v}}}$ is generated based on distribution \eqref{v_h} as the following:
\begin{align}
{\bf{\hat v}} \sim p\left( {{\bf{v}}\left| {{\bf{\hat h}}} \right.} \right).
\end{align}

Furthermore,  the variables ${\bf{W}},{\bf{a}},{\bf{b}}$ are respectively updated as the following rules:
\begin{subequations}
\begin{align}
{\bf{W}} &\leftarrow {\bf{W}} + \varepsilon \Delta {\bf{W}},\hfill \\
{\bf{a}} &\leftarrow {\bf{a}} + \varepsilon \Delta {\bf{a}},\hfill \\
{\bf{b}} &\leftarrow {\bf{b}} + \varepsilon \Delta {\bf{b}}, \hfill
\end{align}
\end{subequations}
where $\varepsilon$ denotes  the learning rate, $\Delta {\bf{W}}$, $\Delta {\bf{a}}$ and $\Delta {\bf{b}}$ are the partial  gradients of the objective function with respect to ${\bf{W}}$, ${\bf{a}}$ and ${\bf{b}}$, respectively, as calculated in \eqref{derivation}.
By exploiting
the gradient descent method, we obtain
  the optimal parameters ${{\bf{\hat W}}}$, ${{\bf{\hat a}}}$ and ${{\bf{\hat b}}}$ for the first RBM.

  Then, the hidden layer ${\bf{h}}$ of the first RBM can be viewed as the visible layer of the second RBM, whose  hidden layer  is denoted as   ${{\bf{h}}^{\left( 1 \right)}}$.
After training the weight matrix and bias of the second RBM,     ${{\bf{h}}^{\left( 1 \right)}}$ and ${{\bf{h}}^{\left( 2 \right)}}$ are viewed as the visible layer and hidden layer of the third RBM, respectively.
After the third RBM is trained, all the parameters (weights and biases) of the RBMs are fine-tuned by a
supervised back-propagation (BP) algorithm \cite{BackPropagation}.
After training, the parameters of the DBN model are
updated  to approach the optimal classifier.
The DBN  is applied to demodulate signals at the test phase, where the demodulation results $\hat s$ is corresponding to classification result $\hat z$.

\section{AdaBoost based Demodulator}
\begin{figure}[h]
          \centering
      \includegraphics[width=8cm]{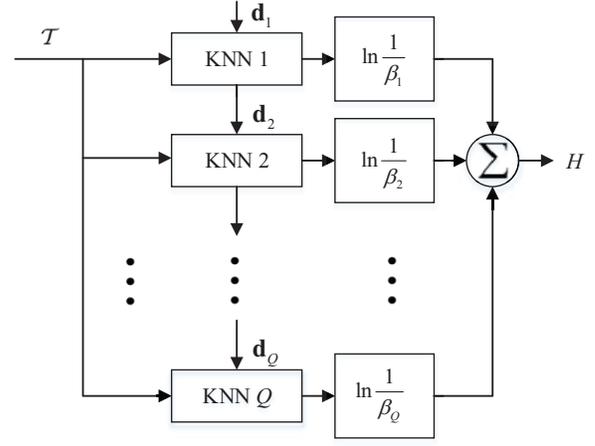}
 \caption{~The generation process of the strong classifier.}
  \label{adaboost} 
\end{figure}

AdaBoost algorithm is a powerful tool that can integrate multiple independent weakly classifiers into a high-performance stronger classifier.
In this paper, we exploit the AdaBoost  method to demodulate signals, where the generation process of strong classifier  is shown in Fig. \ref{adaboost}.
Here, KNN is employed as the weak classifier.

Suppose that the strong classifier is composed of $Q$ KNNs \cite{Convergence}.
 For the $q$th KNN, the weight of samples in ${\rm{{\cal T}}}$ is represented by
${{\bf{d}}_q} = {\left[ {{d_{q,1}},{d_{q,2}},...,{d_{q,K}}} \right]^{\rm{T}}}$, $q = 1,2,...,Q$, and ${d_{q,i}}$ stands for the weight of the $i$th sample in ${\rm{{\cal T}}}$.
When $q = 1$, ${d_{q,i}} = 1/K$, $i = 1,2,...,K$.
The training set of the $q$th KNN is represented by ${{\rm{{\cal T}}}_q}$, which is generated by re-sampling of ${\rm{{\cal T}}}$ according to ${{\bf{d}}_q}$ \cite{Margins}.
Assume that ${{\rm{{\cal T}}}_q} = \left\{ {\left( {{{\bf{x}}_{q,1}},{z_{q,1}}} \right),\left( {{{\bf{x}}_{q,2}},{z_{q,2}}} \right),...,\left( {{{\bf{x}}_{q,K}},{z_{q,K}}} \right)} \right\}$,
and $\left( {{{\bf{x}}_{q,i}},{z_{q,i}}} \right) \in {\rm{{\cal T}}}$.
The testing set is ${\rm{{\cal T}}}$.
${{{\bf{\tilde y}}}_i}$ stands for  the nearest sample of ${{{{\bf{\bar y}}}_i}}$ in training set ${{\rm{{\cal T}}}_q}$, i.e.,
\begin{align}
{{{\bf{\tilde y}}}_i} = \arg \mathop {\min }\limits_{\left\{ {{{\bf{x}}_{q,i}}} \right\}_{i = 1}^K} {\left\| {{{\bf{x}}_{q,i}} - {{{\bf{\bar y}}}_i}} \right\|_2},
\end{align}
where ${\left\| {{{\bf{x}}_{q,i}} - {{{\bf{\bar y}}}_i}} \right\|_2}$ is the Euclidean distance between ${{{\bf{x}}_{q,i}}}$ and ${{{{\bf{\bar y}}}_i}}$.
Assume that the label of  ${{{\bf{\tilde y}}}_i}$ is ${{\tilde z}_i}$, the KNN classifier categorizes ${{{{\mathbf{\bar y}}}_i}}$ to ${{\tilde z}_i}$.
Hence, the classifier can be represented by ${G_q}\left( {{{{\mathbf{\bar y}}}_i}} \right) = {{\tilde z}_i}$, which means
the classification result of the $q$th KNN for sample ${{{{\mathbf{\bar y}}}_i}}$ is ${{\tilde z}_i}$.

The error of ${G_q}$ is defined as weighted sum of weights of the misclassified samples \cite{adaboost}:
\begin{align}
{e_q} = \sum\limits_{i = 1}^K {{d_{q,i}}\left( {1 - I\left( {{G_q}\left( {{{{\bf{\bar y}}}_i}} \right),{z_i}} \right)} \right)} ,
\end{align}
where $I\left( {a,b} \right)$ is indication function: $$I\left( {a,b} \right) = \left\{ {\begin{array}{*{20}{c}}
{1,\;\;\;{\mkern 1mu} {\kern 1pt} {\rm{if}}\quad a = b,}\\
{0,\;\;\;{\mkern 1mu} {\kern 1pt} {\rm{if}}\quad a \ne b.}
\end{array}} \right.$$

Similarly, let ${{\mathbf{d}}_{q + 1}} = {\left[ {{d_{q + 1,1}},{d_{q + 1,2}},...,{d_{q + 1,K}}} \right]^{\text{T}}}$ stand for the weight of samples for the $q+1$th KNN,
and it can be obtained by:
\begin{align}
{d_{q + 1,i}} = {d_{q,i}}{e^{\ln \frac{1}{{{\beta _q}}}\left( {1 - I\left( {{G_q}\left( {{{{\bf{\bar y}}}_i}} \right),{z_i}} \right)} \right)}},i = 1,2,...,K,
\end{align}
where ${\beta _q}$ is computed as a function of ${{e_q}}$ such that ${\beta _q} = \frac{{{e_q}}}{{1 - {e_q}}}$.
Under the constraints of ${e_q} < 0.5$, ${\beta _q} < 1$.
If ${{{{\bf{\bar y}}}_i}}$ is correctly classified, we have $I\left( {{G_q}\left( {{{{\bf{\bar y}}}_i}} \right),{z_i}} \right) = 1$,
${d_{q + 1,i}} = {d_{q,i}}$.
If ${{{{\bf{\bar y}}}_i}}$ is misclassified, $I\left( {{G_q}\left( {{{{\bf{\bar y}}}_i}} \right),{z_i}} \right) = 0$,
and   ${d_{q + 1,i}}\left( i \right) = \frac{{{d_{q,i}}}}{{{\beta _q}}}$.

We redefine ${d_{q + 1,i}}$ by the following normalization formula:
\begin{align}
{d_{q + 1,i}} = \frac{{{d_{q + 1,i}}}}{{\sum\limits_{k = 1}^K {{d_{q + 1,k}}} }}.
\end{align}

After generating $Q$ KNNs, the strong classifier is determined by:
\begin{align}
H\left( {\mathbf{y}} \right) = \hat z = \arg \mathop {\max }\limits_{z \in C} \sum\limits_{q = 1}^Q {\ln \frac{1}{{{\beta _q}}}I\left( {{G_q}\left( {\mathbf{y}} \right),z} \right)},
\end{align}
where ${\bf{y}}$ denotes the  test sample, ${\ln \frac{1}{{{\beta _q}}}}$ is the coefficient of ${{G_q}}$.
${I\left( {{G_q}\left( {\bf{y}} \right),z} \right)}$ can be treated as the voting value, i.e.:
if $I\left( {{G_q}\left( {\bf{y}} \right),z} \right) = 1$, ${{G_q}}$ classifies sample ${\bf{y}}$ into class $z$, otherwise ${\bf{y}}$ does not belong to class $z$.
The class with the maximum sum of weighted voting value $\ln \frac{1}{{{\beta _q}}}I\left( {{G_q}\left( {\bf{y}} \right),z} \right)$ for all classifiers is identified as
the classification result ${\hat z}$ of the AdaBoost classifier, and then ${\hat z}$ is mapped to demodulation result $\hat s$.

\section{Experiment Results and Discussions}

\subsection{The End-to-End VLC   System Prototype}

\begin{figure}[h]
          \centering
      \includegraphics[width=9cm]{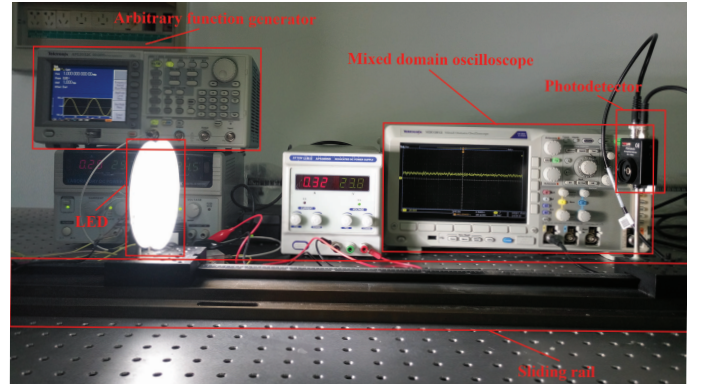}
 \caption{~The devices of the end-to-end VLC   system prototype.}
  \label{equb} 
\end{figure}

As shown in Fig. \ref{equb}, the  proposed end-to-end VLC system prototype includes a source computer,  an arbitrary function generator,
 an amplifier, a bias-T, a LED, a sliding rail, a PD,
and a mixed domain oscilloscope.  We use this prototype to generate the  real VLC modulation  dataset   and verify the proposed data-driven demodulation methods.
The  parameters of the devices used in the end-to-end VLC system
 prototype are listed in Table \ref{Equipments1}.

In the experiments, a serial binary bit stream is randomly generated and modulated in $8$ different types of signals on computer with MATLAB.
We sample $N$ points in one period to generate modulated digital signals for each scheme, which is transferred to analog waveforms by the arbitrary function generator.
The modulated current after amplification is superimposed on LED.
At the receiver, the sampled digital signals are monitored and shown by the mixed domain oscilloscope.
After normalization, we treat the signal in one period as input of DBN for training and testing.
In CNN, we transfer the vector into image as demonstrated in Section III.
The vector is considered as the feature of transmitted symbol and processed by  AdaBoost so that it can be demodulated.

\begin{table}[htbp]
\centering
\caption{Devices and parameters of the VLC system prototype}
\label{Equipments1}
\centering
\begin{tabular}{ll}
\toprule
Device/Parameter  & value  \\
\midrule
Arbitrary function generator  & Tektronix AFG3152C  \\
Sampling rate & 2500000 samples/second\\
Amplifier & Mini-Circuits ZHL-6A-S+ \\
Gain of amplifier  &  25dB \\
 Bias-T & SHWBT-006000-SFFF \\
PD  & PDA10A-EC  \\
Field of view (FOV) of PD &  ${90^\circ }$ \\
Responsivity of PD & 0.44A/W at 750nm \\
Mixed domain oscilloscope  & Tektronix MDO3012  \\
Power of LED  &  7.35W\protect\footnote{\protect{The voltage of the LED in our experiment is $30$V, and the current is about $0.245$A.}}\\
Half-intensity radiation angle & ${60^\circ }$ \\
\bottomrule
\end{tabular}
\end{table}

Our open dataset\footnote{The dataset is collected in real physical environment, and the channel suffers from many factors such as limited LED bandwidth, multi-reflection, spurious or continuous jamming, etc.} contains eight modulation   types, i.e.,  OOK, QPSK, $4$-PPM, $16$-QAM, $32$-QAM, $64$-QAM, $128$-QAM and $256$-QAM.
 For each type of
  modulation, there are
 four different numbers of sample points in each period, i.e., $N=10,20,40,80$.
 The number of periods in each case is listed in Table \ref{dataset1}. Specially, $N=8,16,32,64$ for $4$-PPM.
 \begin{table}[htbp]
\caption{The structure and size of the dataset}
\label{dataset1}
\centering
\begin{tabular}{|c|c|c|c|c|}
\hline
\multirow{2}{*}{Modulation}& \multicolumn{4}{c|}{ $N$ } \\
\cline{2-5} & 10 & 20 & 40& 80 \\
\hline
 OOK & 72000     & 36000 & 18000 & 18000 \\
 QPSK&  72000     & 36000 & 18000 & 18000 \\
 $4$-PPM &   90000     & 45000 & 22500 & 11250\\
 $16$-QAM &   67500    & 33750 & 18000 & 18000 \\
 $32$-QAM &   81000    & 36000 & 36000 & 36000\\
 $64$-QAM &   81000     & 72000 & 72000 & 72000\\
 $128$-QAM &   81000    & 72000 & 72000& 72000\\
 $256$-QAM &   81000     & 72000& 72000& 72000\\
\hline
\end{tabular}
\end{table}
 Let $d$ denote the distance between  LED and  PD.
  The data is collected for every $5$cm from $d=0$cm to $d=140$cm
  and  normalized.
 The illuminance of the  ambient light  is about $85$ Lux. At the distance of $d=100$cm,  the illuminance of the LED is $492$ Lux.
Our database is available at https://pan.baidu.com/s/1rS143bEDaOTEiCneXE67dg.
Eight modulation schemes are tested  in   experiments, where the numbers of signal periods   for training and testing are listed in Table \ref{amounts}.  For the DBN demodulator, we adopt the gradient  descent method in pre-training stage. Then,
the parameters are fine-tuned by the BP algorithm \cite{BackPropagation}. For the CNN demodulator,  the BP algorithm is also used  to   train parameters.

\begin{table}[htbp]
\caption{Training and testing data set}
\label{amounts}
\centering
\begin{tabular}{|c|c|c|}
\hline
\multirow{2}{*}{Modulation} & \multicolumn{2}{c|}{Number of signal periods} \\
\cline{2-3} & Training & Testing \\
\hline
 OOK & \quad 12000 \quad \quad & 6000 \\
 QPSK& \quad 12000 \quad \quad & 6000 \\
 $4$-PPM & \quad 7500 \quad \quad & 3750 \\
 $16$-QAM & \quad 12000 \quad \quad& 6000 \\
 $32$-QAM & \quad 24000 \quad \quad& 12000\\
 $64$-QAM & \quad 48000 \quad \quad & 24000\\
 $128$-QAM & \quad 48000 \quad \quad& 24000\\
 $256$-QAM & \quad 48000 \quad \quad & 24000\\
\hline
\end{tabular}
\end{table}

\subsection{Experiment Results}

The DBN used in the experiments consists of $10$, $20$, $40$ and $80$ visible units according to the dimension of the input data, and the size of output layer is determined by
the demodulation scheme used.
There are three hidden layers, and the size of each hidden layer and training parameters are listed in Table \ref{DBNtab}.
For OOK signals, the three hidden layers  have $10$, $10$, and $20$ hidden units respectively.
For $256$-QAM signals, there are $500$, $500$, and $2000$ hidden units of the three layers.
As for the CNN based demodulator, the batch size is $100$ and the epoch number is $100 \sim 200$.
\begin{table}[htbp]
\caption{DBN structure and parameters.}
\centering
\begin{tabular}{ll}
\toprule
Size of hidden layer-1 & $10 \sim 500$ \\
Size of hidden layer-2 & $10 \sim 500$ \\
Size of hidden layer-3 & $20 \sim 2000$ \\
Pre-training epoch & $50 \sim 1000$ \\
BP epoch & $50 \sim 1000$ \\
Batch size & 100 \\
Learning rate & 0.1\\
\bottomrule
\end{tabular}
\label{DBNtab}
\end{table}

All the proposed methods are implemented with MATLAB R2016b and executed on a computer with an Intel Core i7-7700 CPU @ 3.60 GHz/32 GB RAM.
The DeepLearnToolbox \cite{IMM2012-06284} is used to implement the CNN and DBN based classifiers.
We first investigate the performance of the proposed CNN, DBN, and  AdaBoost based  demodulation methods versus distance $d$ with $N= 40$.
 After training, the accuracies on test set is calculated.
 Moreover, both the support vector machine (SVM) based
 and  the maximum likelihood (MLD) based demodulation methods are used for comparison.
SVM is a supervised learning method which solve binary classification problems. In this paper, we combine SVMs to demodulate by one-to-one way.
MLD classification is one of the supervised classification algorithms based on the Bayesian criterion,
which assumes that the input feature vector follows  $N$-dimensional normal distribution, and calculate the attribution probability of the input vector belonging to each category.
The data vector is categorized to the class with the  maximum attribution probability.

\begin{figure}
        \begin{minipage}[b]{0.45\textwidth}
      \centering
      \includegraphics[height=7cm,width=9cm]{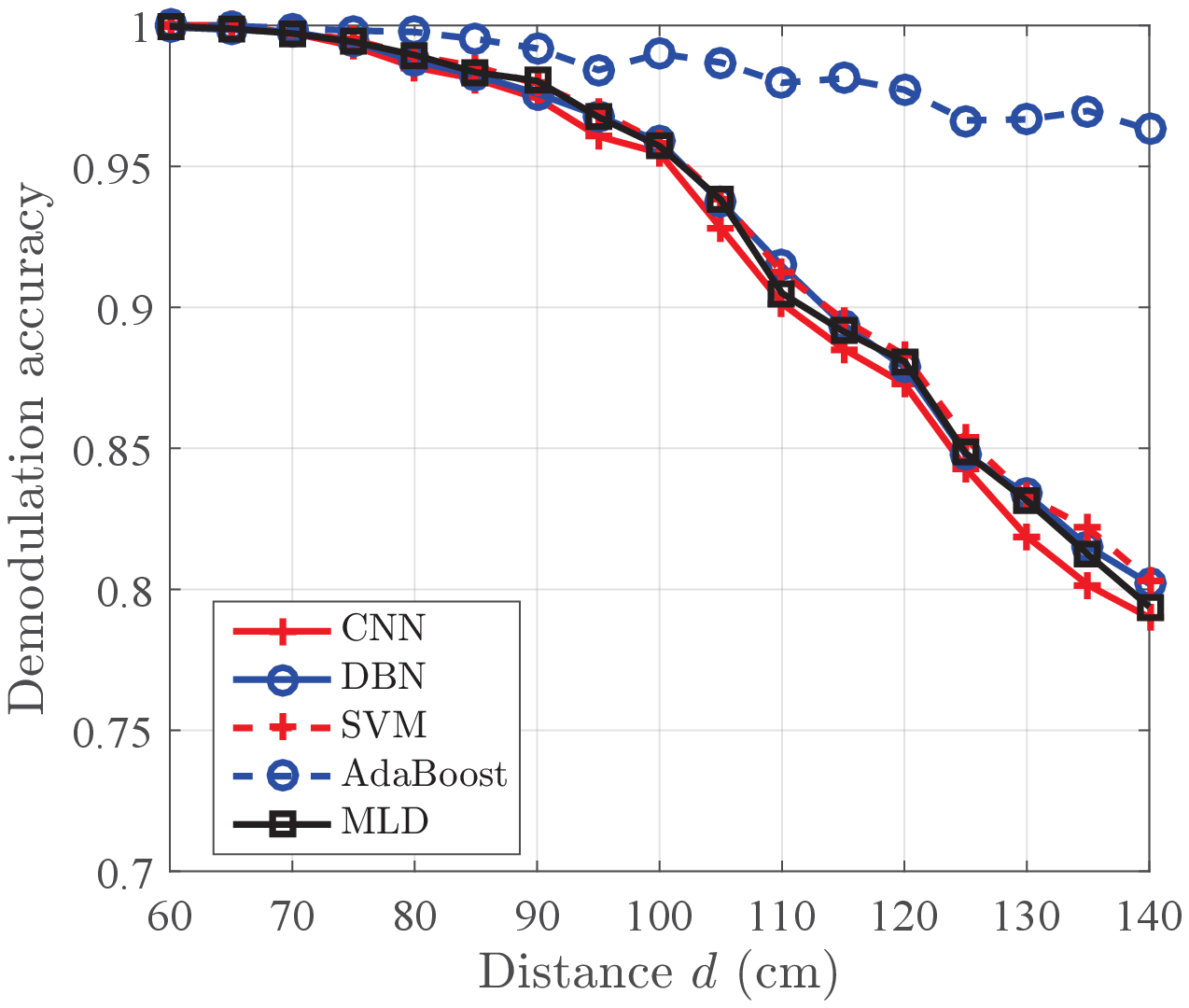}
      \vskip-0.2cm\centering {\footnotesize (a)}
    \end{minipage}
            \begin{minipage}[b]{0.45\textwidth}
      \centering
      \includegraphics[height=7cm,width=9cm]{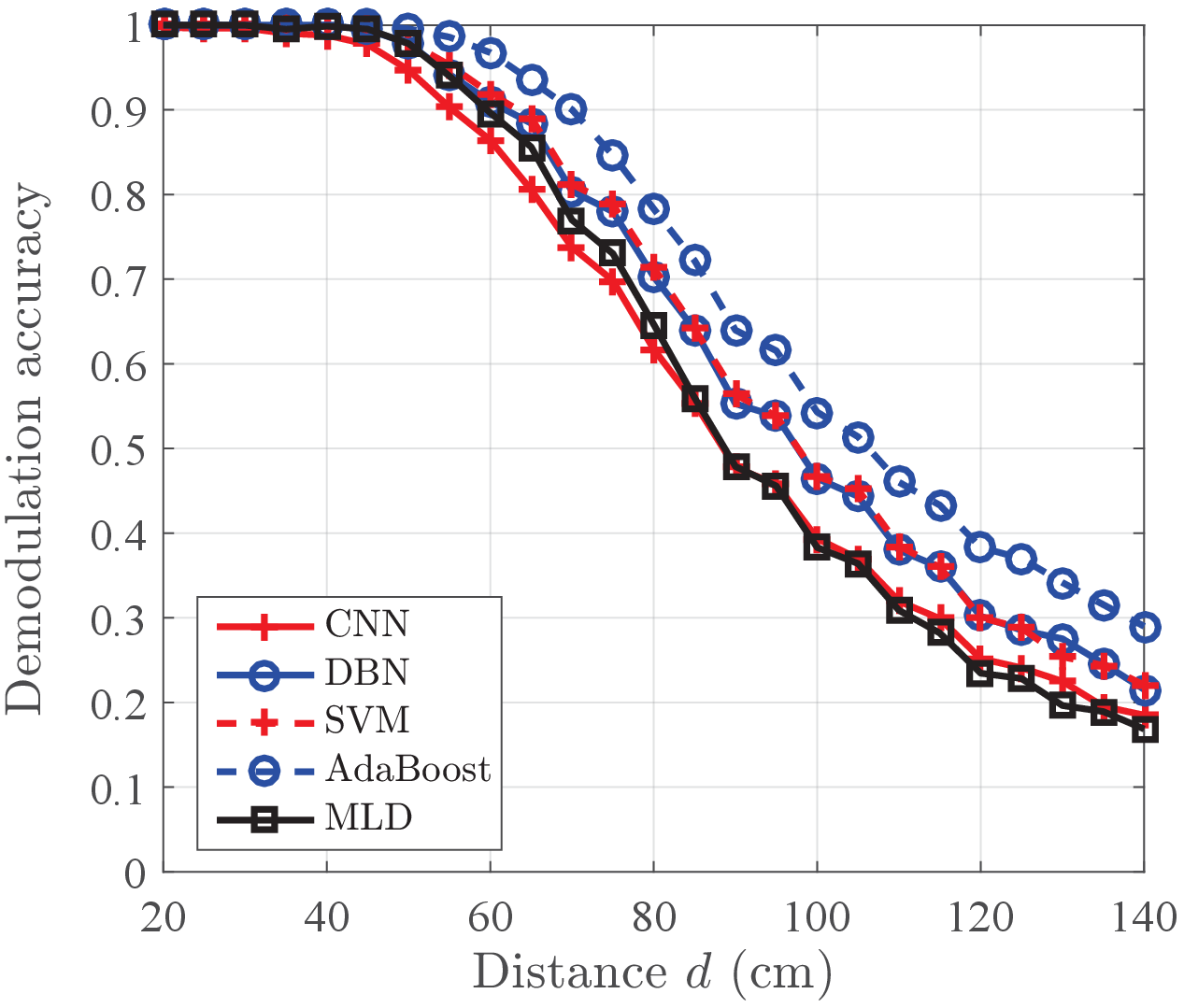}
      \vskip-0.2cm\centering {\footnotesize (b)}
    \end{minipage}
    \begin{minipage}[b]{0.45\textwidth}
      \centering
      \includegraphics[height=7cm,width=9cm]{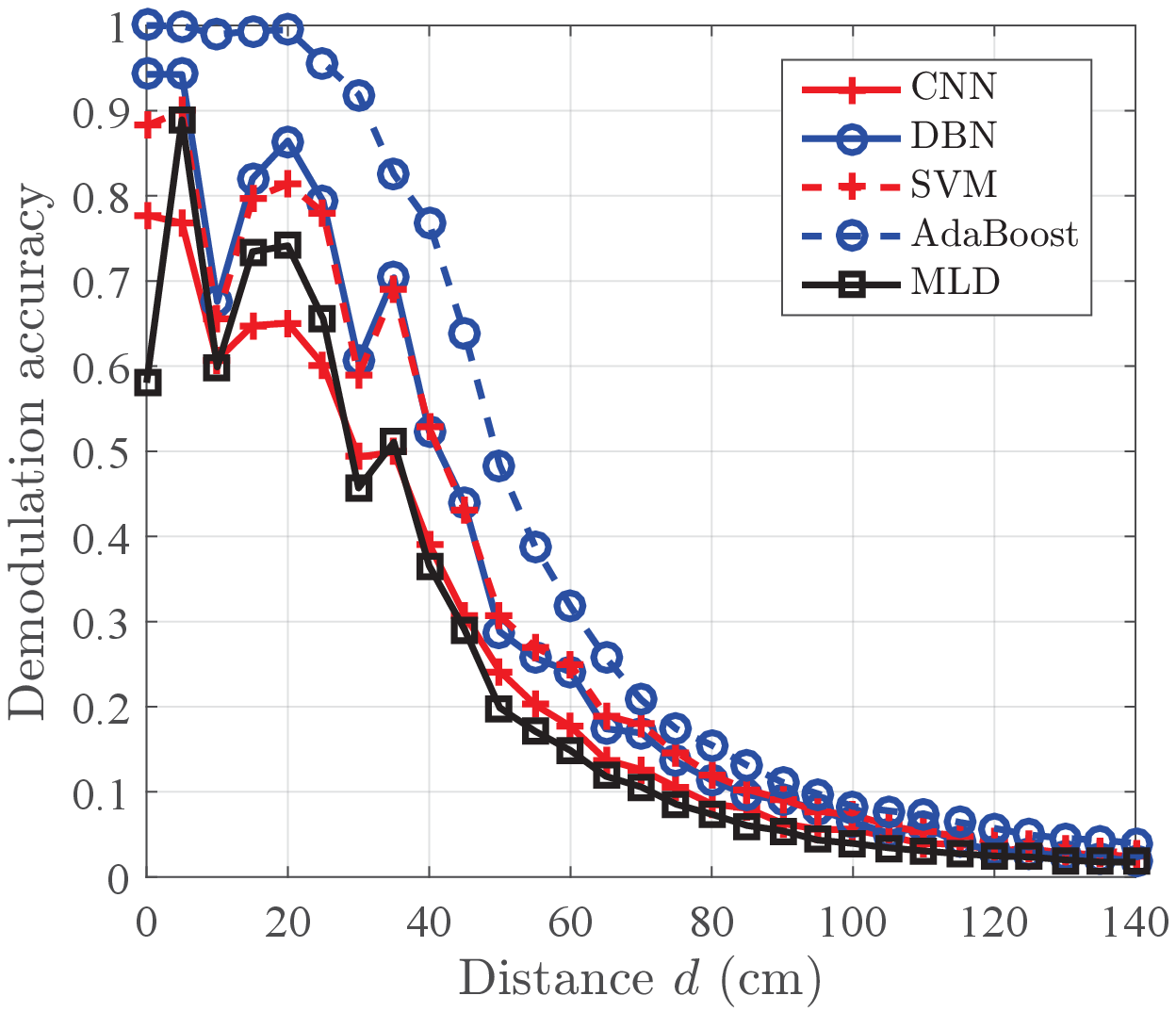}
      \vskip-0.2cm\centering {\footnotesize (c)}
    \end{minipage}\hfill
 \caption{(a)~The demodulation accuracy of  OOK modulated signals versus  distance $d$ when $N= 40$;
   (b)~The demodulation accuracy of 32-QAM modulated signals versus  distance $d$ when $N= 40$;
 (c)~The demodulation accuracy of 256-QAM modulated signals versus  distance $d$ when $N= 40$.}
  \label{demodulation_types} 
\end{figure}
Fig. \ref{demodulation_types} (a), (b) and (c) show the
demodulation accuracies of symbols of  OOK, $32$-QAM and $256$-QAM  modulated signals versus  distance $d$, respectively.
We can see that the demodulation accuracy of all methods  decreases  as the  distance $d$ increases.
Specifically, Fig. \ref{demodulation_types} (a)  shows that the  demodulation accuracy of all methods of  OOK   modulated signals
are close to $100\%$ for $d \le 70$cm; and for $70$cm $< d \le 140$cm, the
 proposed AdaBoost based  demodulation method  significantly
outperforms  other demodulation methods.
 Fig. \ref{demodulation_types} (b) shows that the  demodulation accuracies of the  32-QAM   modulated signals by all methods
are close to $100\%$ for $d \le 40$cm. For $40$cm $< d \le 140$cm,
the demodulation accuracy of the AdaBoost based  demodulation method is the highest among the five demodulation methods.
The accuracies of the DBN
 and SVM based demodulation methods are  similar, but higher than that of both CNN  and MLD based demodulation methods.
The reason might be referring to the fact that CNNs ignore the classical sampling theorem, so that the performance cannot be guaranteed \cite{azulay2018deep}.
Besides, the combined output after down sampling is typically the scalar activity of the most active unit in the pool \cite{riesenhuber},
and the relative position information of parts of waveforms is ignored.
Since the practical VLC channels include complex interferences, the MLD classification have a degraded performance.

In  Fig. \ref{demodulation_types} (c) it is shown that for the 256-QAM  modulated signals, the  demodulation accuracies of all mothods are similar to Fig. \ref{demodulation_types} (b).


\begin{figure}[h]
          \centering
      \includegraphics[height=7cm,width=9cm]{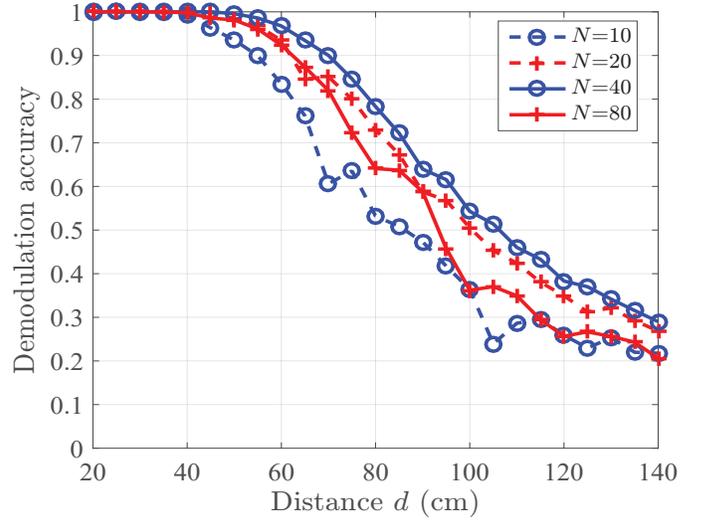}
 \caption{~The demodulation accuracy of AdaBoost versus distance $d$.}
  \label{32qam-AdaBoost} 
\end{figure}
Fig. \ref{32qam-AdaBoost} shows the
 accuracy  of AdaBoost based demodulation method  versus  distance $d$ with different numbers of sample points in one period $N= 10, 20, 40, 80$,
 where   the signals are modulated by 32-QAM.
The demodulation accuracy increases as  number of sample points $N$ increases.
Moreover, the demodulation accuracy of the $N=40$ case is higher than that of
$N=80$ case, while the storage memory of the $N=40$ case is only a half of that of the $N=80$ case.

\begin{figure}[h]
          \centering
      \includegraphics[height=7cm,width=9cm]{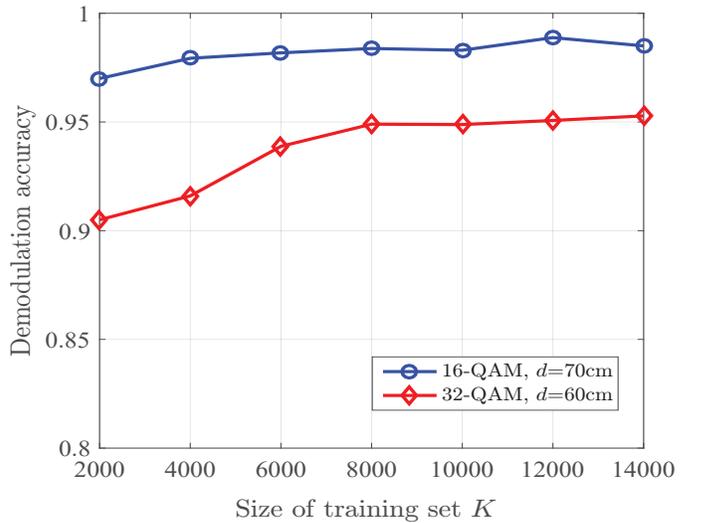}
 \caption{~The demodulation accuracy of 16-QAM and 32-QAM modulated signals versus   number of training periods $K$.}
  \label{16qam-32qam_amount} 
\end{figure}

Fig. \ref{16qam-32qam_amount} shows the demodulation accuracy of the AdaBoost demodulation method versus
the number of training periods $K$ with $16$-QAM modulated signals at $d = 70$cm and $32$-QAM modulated signals at $d = 60$cm.
For the $16$-QAM modulated signal case, the demodulation accuracy  increases with the number of training periods $K$, while when  $K \ge 4000$,
the demodulation accuracy increases very slowly.
Similarly, for the $32$-QAM modulated signal case, the demodulation accuracy  increases with the number of training periods $K$,
and when  $K\ge 8000$, the demodulation accuracy almost keeps the same.
Comparing demodulation accuracy of the $16$-QAM and $32$-QAM modulated signals, it can be observed that more  number of training periods $K$ is required for the higher modulation order
to achieve a stable accuracy.

\begin{figure}
        \begin{minipage}[b]{0.45\textwidth}
      \centering
      \includegraphics[height=7cm,width=9cm]{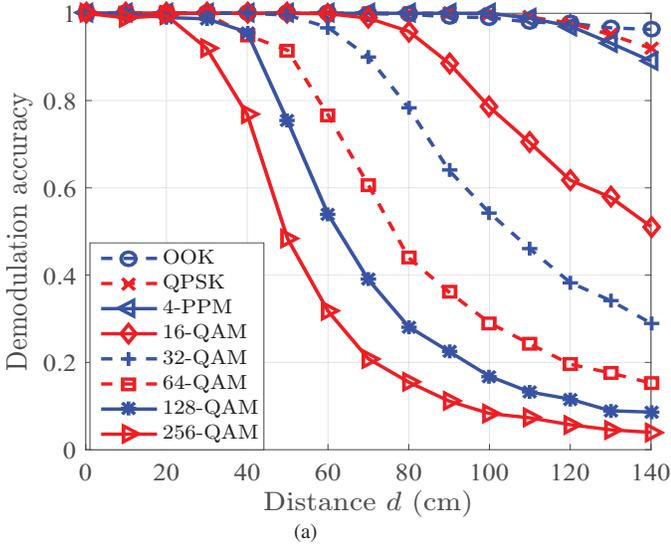}
      \vskip-0.2cm\centering {\footnotesize (a)}
    \end{minipage}
            \begin{minipage}[b]{0.45\textwidth}
      \centering
      \includegraphics[height=7cm,width=8.8cm]{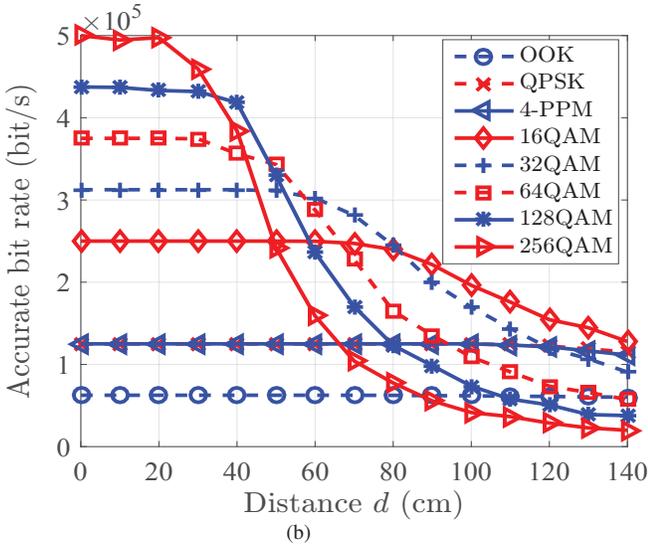}
      \vskip-0.2cm\centering {\footnotesize (b)}
    \end{minipage}
 \caption{(a)~The demodulation   accuracy  of  OOK, QPSK, $4$-PPM, $32$-QAM, $64$-QAM, $128$-QAM and $256$-QAM  modulated signals versus  distance $d$ when $N=40$;
   (b)~The accurate bit rate of  OOK, QPSK, $4$-PPM, $32$-QAM, $64$-QAM, $128$-QAM and $256$-QAM  modulated signals versus  distance $d$ when $N=40$.}
  \label{Adaboost-40} 
\end{figure}

Fig. \ref{Adaboost-40} (a) shows    the
demodulation accuracies of  OOK, QPSK, $4$-PPM, $32$-QAM, $64$-QAM, $128$-QAM and $256$-QAM  modulated signals versus  distance $d$ when $N=40$, respectively.
We can see that the demodulation accuracies of the eight modulation schemes  decrease  as the  distance $d$ increases.
Moreover, for a given distance $d$, the   demodulation accuracies of the eight modulation schemes
  decrease as the modulation order increases, and the higher the modulation order is, the faster of the rate decreases.

Fig. \ref{Adaboost-40} (b) shows the accurate bit rates\footnote{The accurate bit rate is the product of demodulate accuracy and the information each symbol carries.} of the eight modulation schemes versus distance $d$.
As distance $d$ increases, the effective rates of the eight modulation schemes decrease.
When $d \le 30$cm, the effective rate   of $256$-QAM is the highest.
When $40$ cm $< d \le 50$cm, the highest  effective rate is obtained by the $128$-QAM modulation scheme.
As distance $d$ increases from $50$cm to $140$cm, the highest accurate bit rate is obtained with $64$-QAM, $32$-QAM and $16$-QAM in turn.
Therefore, for short distance or high SNR scenario, high order modulation is preferred.
\begin{figure}
        \begin{minipage}[b]{0.45\textwidth}
      \centering
      \includegraphics[height=7cm,width=9cm]{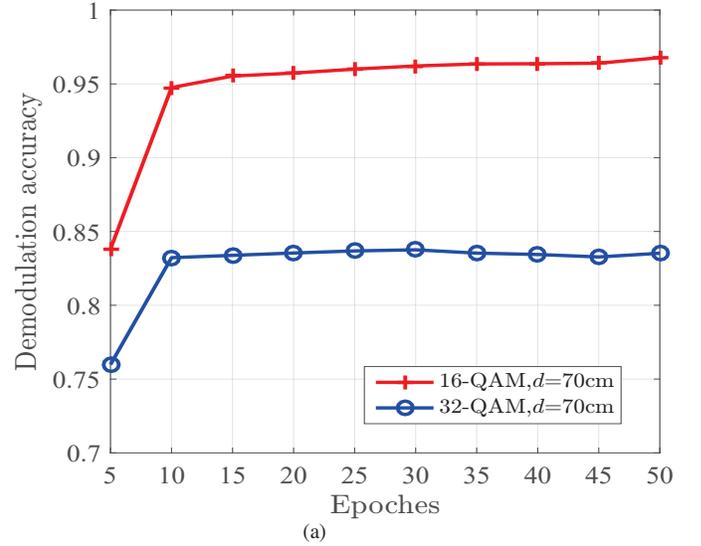}
      \vskip-0.2cm\centering {\footnotesize (a)}
    \end{minipage}
            \begin{minipage}[b]{0.45\textwidth}
      \centering
      \includegraphics[height=7cm,width=9cm]{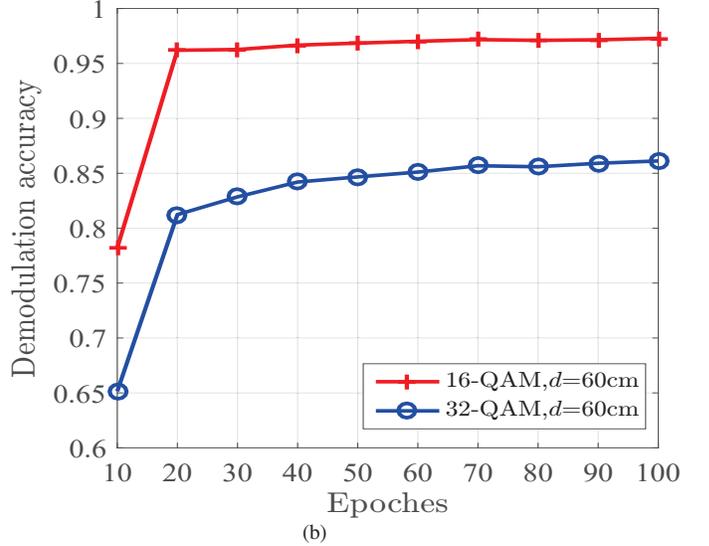}
      \vskip-0.2cm\centering {\footnotesize (b)}
    \end{minipage}
 \caption{(a)~The demodulation  accuracy  of DBN based method versus epoches when $N=40$ and $d=70$cm;
   (b)~The demodulation  accuracy  of CNN based method versus epoches when $N=40$ and $d=60$cm.}
  \label{numepoches} 
\end{figure}

Fig. \ref{numepoches} (a) shows the  demodulation  accuracies of the DBN demodulation method versus epoches in BP process with  $N=40$ and $d=70$cm.
We can see that the demodulation accuracies of $16$-QAM and $32$-QAM modulated signal increase  as the number of epoch increases.
For the two modulation schemes, the demodulation accuracy  increases fast when the number of epoch is less than $10$, while the performance of $16$-QAM is higher than $32$-QAM.
When larger than $10$, the increasing of epoch number brings limited benefits.
Fig. \ref{numepoches} (b) shows the  demodulation  accuracies of the CNN demodulation method versus epoches when $N=40$ and $d=60$cm.
The  demodulation accuracies of two modulation schemes are similar to Fig. \ref{numepoches} (a).

\section{Conclusion}


In this paper,  three  data-driven demodulators (CNN, DBN, and AdaBoost) based demodulators are designed for the physical layer of VLC systems.
A flexible end-to-end VLC system
prototype is constructed for real data collection. By using the proposed prototype,   an open online real modulated  dataset is
created, which consists   eight types of modulated  signals, i.e.,  OOK, QPSK, $4$-PPM, $16$-QAM,
$32$-QAM, $64$-QAM, $128$-QAM and $256$-QAM.
  Based on this real dataset,  we investigate the demodulation performance of the proposed  three demodulators.
     Experimental results show that for a given transmission   distance, the demodulation accuracy decreases as the modulation order increases.
Moreover, that the demodulation accuracy of the AdaBoost based demodulators is higher than other demodulators.
For the short distance or high SNR scenario, a high-order modulation is preferred.
In the future, we  will further investigate   dedicated ML based demodulators for VLC systems.

\bibliographystyle{IEEE-unsorted}
\bibliography{VLC_0303}
\end{document}